\documentstyle[12pt]{article}
\begin{document}
\baselineskip 18pt
\begin{center}
{\Large Baryogenesis Through Mixing of Heavy Majorana Neutrinos}\\
\vskip .3in
{\large Marion Flanz, Emmanuel A. Paschos, 
Utpal Sarkar\footnote{ {\it Permanent address :} Theory Group, 
Physical Research Laboratory, \\ Ahmedabad-380009, India} }\\  
and  \\
{\large Jan Weiss } \\ 
\vspace{1cm}
{\large Institut f\"{u}r Physik }\\ {\large Universit\"{a}t Dortmund }\\
{\large D-44221 Dortmund, Germany}\\

\end{center}

\vskip  .75in  
\begin{abstract}  
\baselineskip  18pt 

A mechanism  is  presented,  in which the mixing of right  handed
heavy  Majorana  neutrinos  creates a  $CP-$asymmetric  universe.
When these  Majorana  neutrinos  subsequently  decay more leptons
than anti-leptons are produced.  The lepton asymmetry  created by
this new  mechanism  can exceed by a few orders of magnitude  any
lepton  asymmetry  originating from direct decays.  The asymmetry
is  finally   converted  into  a  baryon  asymmetry   during  the
electroweak phase transition.

\end{abstract}

\newpage
\baselineskip 18pt

The  generation of the baryon  asymmetry in the universe has been
discussed  in many  articles \cite{sak,kolb}.  Two  prominant  scenarios  are the
Grand  Unified  Theories  \cite{gut,kolb}  and  the  production  of an
asymmetry   through   extended   solutions   of  field   theories
(sphalerons)  \cite{krs,shap}.  At this time both  scenarios  generate
asymmetries  which are small.  This comes  about  because  in the
former case  $CP-$violation  is produced by higher order  effects
and in the latter the tunneling rate through  potential  barriers
is small (in addition, if the higgs particles are heavier than 80
GeV, then the baryon  asymmetry thus generated will be completely
erased) \cite{shap}.

A third scenario  includes heavy Majorana  neutrinos whose decays
generate a lepton asymmetry, which later on is converted into the
baryon asymmetry.  By their very nature Majorana neutrinos posses
$\Delta  L =  2$  transitions  and  in  addition  they  may  have
couplings  which allow them to decay into the standard  higgs and
leptons, i.e., $N \rightarrow  \phi^\dagger l^-$  \cite{fy}.
In these  models the  $CP-$violation  is  introduced  through the
interference of tree-level  with one-loop  diagrams in the decays
of heavy neutrinos, which we shall call  $\epsilon^\prime  -$type
effects  (direct  $CP-$violation) \cite{fy,covi}. An additional 
contribution from self energies was included in ref [8-10]. 
A new aspect was pointed  out
\cite{paschos}  when it was  realised  that  the  heavy  physical
neutrino  states  are not $CP-$ or  lepton-number--  eigenstates.
Therefore  as soon as the  physical  states are  formed  there is
imprinted on them a $CP-$asymmetry and a lepton--asymmetry.  This
we shall call $\epsilon -$type effects (indirect $CP-$violation).
It is a  property  which  appears  in   the  eigenstates  of  the
Hamiltonian    at  an   early   epoch   of   thermodynamic
equillibrium,  even before the  temperature of the universe falls
down to the masses of the heavy Majorana particles.

We  demonstrate  the  phenomenon  with  a  Gedanken   experiment.
Consider a universe  consisting of a large $p - \bar{p}$ collider
which  produces $s-$ and  $\bar{s}-$pairs.  The $s$ and $\bar{s}$ quarks
hadronize  into  $K^\circ$  and  $\overline{K^\circ}$ mesons whose
superpositions are the physical states
$$  K_{ {L},{S} } \propto  [ (1 + \epsilon) K^\circ \pm
(1 - \epsilon) \overline{K^\circ} ] . $$
The probability of finding a $|K^\circ >$ is proportional  to $|1
+ \epsilon|^2$  and the probability for a  $|\overline{K^\circ}>$
is proportional to $|1 - \epsilon|^2$  which are not equal.  When
the particles decay
$$ K_L \rightarrow \pi^\pm e^\mp \stackrel{(-)}{\nu}  \quad 
{\rm and } \quad K_S 
\rightarrow \pi^\pm e^\mp \stackrel{(-)}{\nu} $$
the above asymmetry  survives as an asymmetry of the detected  $e^+$'s
and $e^-$'s.  The  electron-positron  asymmetry  generated is the
same for the $K_{L}$ and $K_{S}$ \cite{paschos2}.

In this article we point out that a similar  situation  arises in
the  formation  and decays of  Majorana  neutrinos.  We adopt the
wave  function  formalism to  calculate  the  eigenstates  of the
Hamiltonian  and  their  decay  rates.  In this  formalism  we can
extend the  region of our  calculation  to the case of very small
mass differences using degenerate  perturbation  theory.  We find
that for small mass  differences  between the two generations the
indirect  $CP-$violation   ($\epsilon-$type)  produces  a  lepton
asymmetry much larger than that of the $\epsilon^\prime-$type.

We work in an  extension of the  standard  model where we include
one heavy right handed  Majorana  field per  generation  of light
leptons  ($N_i$,  $i=1,2,3$).  These new fields are singlets with
respect to the standard model.  The lagrangian now contains a
Majorana  mass term and the Yukawa  interactions  of these fields
with the light leptons,
\begin{eqnarray}
{\cal L}_{int} & = & \sum_{i} M_i [\overline{(N_{Ri})^c} N_{Ri} + 
   \overline{N_{Ri}} (N_{Ri})^c] \nonumber \\
 & & + \sum_{\alpha, i} \, h^\ast_{\alpha i} \, \overline{N_{Ri}} 
 \, \phi^{\dagger} \,
     \ell_{L \, \alpha} + \sum_{\alpha, i} h_{\alpha i} \, 
      \overline{\ell_{L \, \alpha}} \phi \, N_{Ri} \\ 
 & & + \sum_{\alpha, i} \, h^\ast_{\alpha i} \, \overline{(\ell_{L 
 \, \alpha}})^c \, \phi \, (N_{Ri})^c + \sum_{\alpha, i} 
 h_{\alpha i} \overline{(N_{Ri})^c} \, 
       \phi^\dagger \, (\ell_{L \, \alpha})^c  \nonumber
\end{eqnarray}
where  $\phi^T =  (-\overline{\phi^\circ},  \phi^-)$ is the higgs
doublet of the  standard  model,  which  breaks  the  electroweak
symmetry  and gives  mass to the  fermions;  $l_L\alpha$  are the
light leptons,  $h_{\alpha i}$ are the complex  Yukawa  couplings
and $\alpha$ is the generation  index.  We have adopted the usual
convention for charge  conjugation :  $N^c = C \overline{N^T}  $.
Without  loss  of  generality  we work in a basis  in  which  the
Majorana mass matrix is real and diagonal with eigenvalues $M_i$.

The states $|N_i>$ decay only into leptons, while the states
$|N^c_i>$ decay only into antileptons (figure 1). 
\begin{figure}[htb]
\mbox{}
\vskip 1.5in\relax\noindent\hskip .2in\relax
\includegraphics{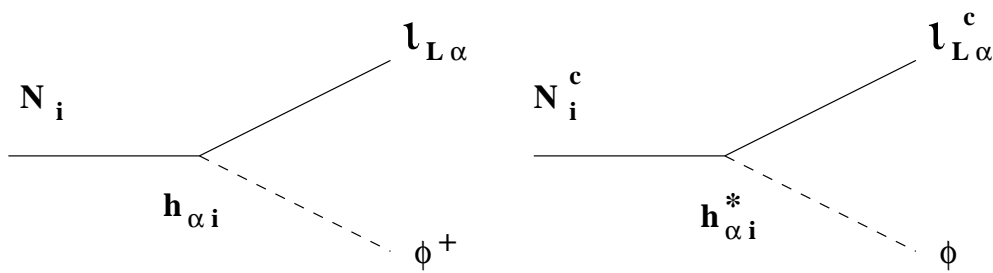} \vskip .25in
\caption{ $N_i$ and $N_i^c$ decaying into $l_{L \alpha}$ and 
$(l_{L \alpha})^c$}
\end{figure}
For this reason the states  $|N_i>$ and $|N_i^c>$  have  definite
lepton  numbers  and  are  the  appropriate  states  to  describe
$CP-$violation in the leptonic sector.  They are analogous to the
$K^\circ$  and  $\overline{K^\circ}$   states.  The  idea  is  as
follows:  Through  the  presence  of the Yukawa  interactions  we
obtain one loop  corrections  to the mass matrix (figure 2), such
that the  corresponding  mass  eigenfunctions  are no longer  the
$|N_i>$ and $|N_i^c>$  states, but a mixture of them. It is these
physical  eigenstates  which  evolve  in  time  with  a  definite
frequency.  If  they  are  shown  to  be  asymmetric   linear
combinations of the $|N_i>$ and $|N_i^c>$'s  then we have created
a $CP-$asymmetric universe.  By asymmetric linear combinations we
mean  that the  $|N_i>$  and  $|N_i^c>$'s  enter  with  different
complex phases into the decomposition of the eigenfunctions.  The
subsequent  decay of these fields will produce the desired lepton
asymmetry.

As a result  even if we start with equal  numbers of $|N_i>$  and
$|N_i^c>$,   they  will  evolve   according  to  the   asymmetric
eigenstates  of definite  time  developement.  Since  $|N_i>$ and
$|N_i^c>$   carry   different   lepton  numbers  given  by  their
interactions,  this means that a lepton  asymmetry is established
through mixing before the fields actually decay.  Herein lies the
main difference between our model and the literature.

For  the  sake of  simplicity  we  consider  two  generations  of
Majorana  neutrinos,  where the  indices $i$ and $j$  take the
values $1$ and $2$.  We assume the hierarchy $M_2 > M_1$.  In the
basis $(|N_1^c> |N_2^c> |N_1> |N_2> )$ the effective  Hamiltonian
of this model can be written as,
\begin{equation}
\widehat{{\cal H}}^{(0)} = \left( \begin{array}{cccccc}
                  0 & 0 & M_1 & 0 \\
                  0 & 0 & 0 & M_2 \\
                  M_1 & 0 & 0 & 0 \\
                  0 & M_2 & 0 & 0 \\
              \end{array}    \right)  .
\end{equation}
Once we include the one loop diagram 
of figure 2, there is an additional 
contribution to the effective Hamiltonian, which introduces $CP-$violation
in the mass matrix. 
\begin{figure}%[htb]
\mbox{}
\vskip 4.25in\relax\noindent\hskip .2in\relax
\includegraphics{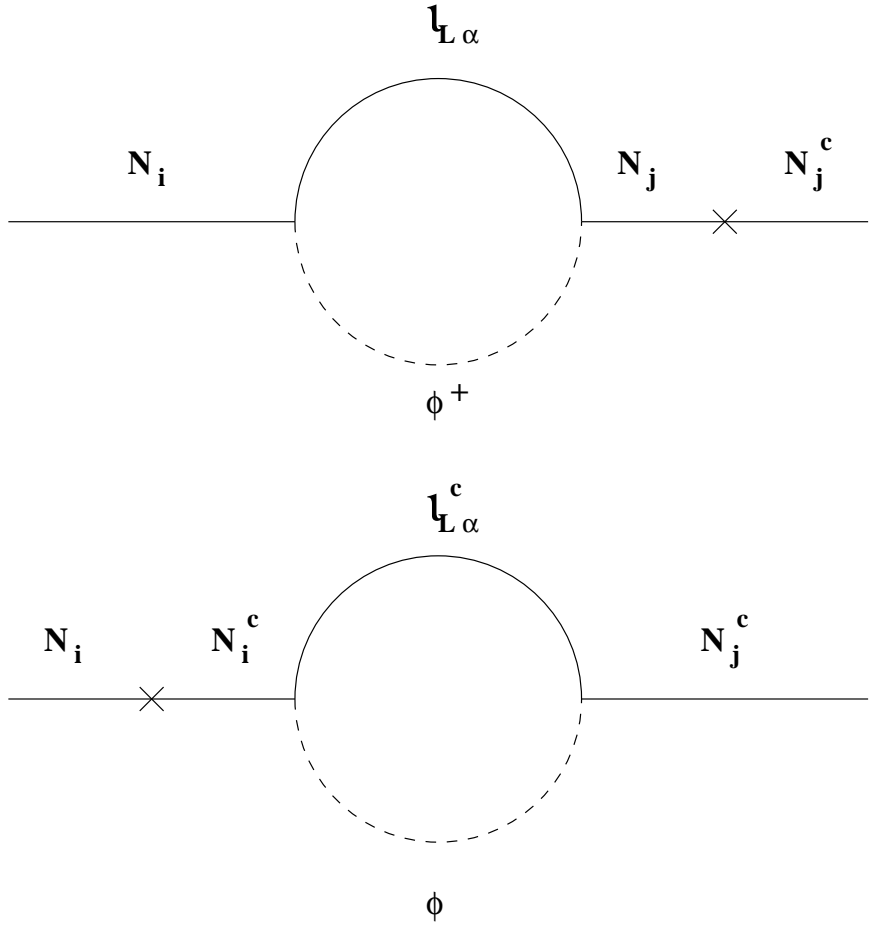}
\caption{ One loop contributions to the mass matrix}
\end{figure}
We treat the one loop contributions as a small perturbation to the
tree level Hamiltonian.  
\begin{equation}
\widehat{{\cal H}}^{(1)} = \left( \begin{array}{cccccc}
                  0 & 0 & H_{11}^{(1)} & H_{12}^{(1)} \\
                  0 & 0 & H_{12}^{(1)} & H_{22}^{(1)} \\
                  H_{11}^{(1)} & \widetilde{H}_{12}^{(1)} & 0 & 0 \\
                  \widetilde{H}_{12}^{(1)} & H_{22}^{(1)} & 0 & 0 \\
              \end{array}    \right)
\end{equation}
with
\begin{equation}
H_{ij}^{(1)} = H_{ji}^{(1)} =
\left[ M_i \sum_\alpha h_{\alpha i}^\ast h_{\alpha j} + 
 M_j \sum_\alpha h_{\alpha i} h_{\alpha j}^\ast \right] (g_{\alpha ij}^{dis}
   - \frac{i}{2} g_{\alpha ij}^{ab})
\end{equation}    
\begin{equation}
\widetilde{H}_{ij}^{(1)} = \widetilde{H}_{ji}^{(1)} =
\left[M_i 
                               \sum_{\alpha} h_{\alpha i} h_{\alpha j}^\ast
     + M_j \sum_{\alpha} h_{\alpha i}^\ast h_{\alpha j} \right] 
     (g_{\alpha ij}^{dis} - \frac{i}{2} g_{\alpha ij}^{ab})
\end{equation}
and 
\begin{equation}
H_{ii}^{(1)} = \widetilde{H}_{ii}^{(1)} =
\left[2 M_i \sum_{\alpha}  h_{\alpha i} h_{\alpha i}^\ast  \right] 
     (g_{\alpha ij}^{dis} - \frac{i}{2} g_{\alpha ij}^{ab})
\end{equation}
as can be easily read off from figure 2. 
The dispersive part $g_{\alpha ij}^{dis}$ 
can be absorbed in the wave function and mass renormalization. 
The absorbtive part $g_{\alpha ij}^{ab}$
of the loop integrals is given by,
\begin{equation}
g_{\alpha ij}^{ab} =  {1 \over 16 \pi }  
\end{equation}
neglecting terms of order $O \left( {m_\alpha^2}/{p^2} \right) $, 
$O \left( {m_\phi^2}/{p^2} \right) $ with $p^2 \geq M_i^2$.

Employing ordinary first order perturbation theory the
eigenfunctions of the effective Hamiltonian $\widehat{\cal H} = 
\widehat{\cal H}^{(0)} + \widehat{\cal H}^{(1)}$ are found to be 
\begin{eqnarray}
|\Psi_1> & = & \frac{1}{\sqrt{\cal N}} (|N_1> + \alpha_2 |N_2> + |N_1^c>  
               + \alpha_1 |N_2^c>) \nonumber \\
|\Psi_1^\prime> & = & \frac{1}{\sqrt{\cal N}} (|N_1> + \alpha_2 |N_2> - |N_1^c>
               - \alpha_1 |N_2^c>) \nonumber \\
|\Psi_2> & = & \frac{1}{\sqrt{\cal N}} (|N_2> - \alpha_1 |N_1> 
               + |N_2^c> - \alpha_2 |N_1^c>) \nonumber \\
|\Psi_2^\prime> & = & \frac{1}{\sqrt{\cal N}} (|N_2> - \alpha_1 |N_1> 
               - |N_2>^c + \alpha_2 |N_1^c>) 
\end{eqnarray} 
with,
\begin{displaymath}
\alpha_1 = \frac{M_1 H^{(1)}_{12} + M_2 
\widetilde{H^{(1)}}_{12}}{M_1^2 - M_2^2} \quad , \quad \quad
\alpha_2 = \frac{M_1 \widetilde{H^{(1)}}_{12} + 
M_2 H^{(1)}_{12}}{M_1^2 - M_2^2}
\end{displaymath}
where the normalization factor is ${\cal N} = 2 + |\alpha_1|^2 
+ |\alpha_2|^2 $.

The states $|\Psi_1>$ and $|\Psi_1^\prime>$ are eigenstates with
mass eigenvalues $\pm (M_1 + H_{11}^{(1)})$. 
They are related by a chiral 
$\gamma_5-$transformation and correspond to the same physical state. 
The same holds for 
the states $|\Psi_2>$ and $|\Psi_2^\prime>$ with eigenvalues 
$\pm (M_2 + H_{22}^{(1)})$.
For the rest of our calculation we 
shall only consider $|\Psi_1>$ and $|\Psi_2>$. 

The asymmetry parameter can be defined as,
\begin{equation}
\Delta = \sum_{i=1}^2 \frac{\Gamma_{\Psi_i \rightarrow l} - 
         \Gamma_{\Psi_i \rightarrow l^c}}{\Gamma_{\Psi_i \rightarrow l} 
         + \Gamma_{\Psi_i \rightarrow l^c}}
\end{equation}
which is a measure of the lepton asymmetry generated when the 
physical states $|\Psi_i>$ finally decay. This can be 
calculated using,

\vbox{
\begin{eqnarray}
\Gamma_{\Psi_1 \rightarrow l} & \propto & \sum_{\alpha} |h_{\alpha 1} + 
              \alpha_2 h_{\alpha 2}|^2 \nonumber \\ 
            & = & \sum_{\alpha} \left[ |h_{\alpha 1}|^2 +
                 |\alpha_2|^2 |h_{\alpha 2}|^2 + 2 {\rm Re}(\alpha_2
                 h_{\alpha 1}^\ast h_{\alpha 2}) \right] \nonumber \\
\Gamma_{\Psi_1 \rightarrow l^c} & \propto &  \sum_{\alpha} |h_{\alpha 1}^\ast 
           + \alpha_1 h_{\alpha 2}^\ast|^2 \nonumber \\
            & = & \sum_{\alpha} \left[ |h_{\alpha 1}|^2 +
                 |\alpha_1|^2 |h_{\alpha 2}|^2 + 2 {\rm Re}(\alpha_1
                 h_{\alpha 1} h_{\alpha 2}^\ast) \right] \nonumber \\
\Gamma_{\Psi_2 \rightarrow l}  & \propto &  \sum_{\alpha} |h_{\alpha 2} 
            - \alpha_1 h_{\alpha 1}|^2 \nonumber \\
             & = & \sum_{\alpha} \left[ |h_{\alpha 2}|^2 +
                 |\alpha_1|^2 |h_{\alpha 1}|^2 - 2 {\rm Re}(\alpha_1
                 h_{\alpha 1} h_{\alpha 2}^\ast) \right] \nonumber \\
\Gamma_{\Psi_2 \rightarrow l^c}  & \propto  & \sum_{\alpha} |h_{\alpha 2}^\ast  
              - \alpha_2 h_{\alpha 1}^\ast|^2 \nonumber \\
             & = & \sum_{\alpha} \left[ |h_{\alpha 2}|^2 +
                 |\alpha_2|^2 |h_{\alpha 1}|^2 - 2 {\rm Re}(\alpha_2
                 h_{\alpha 1}^\ast h_{\alpha 2}) \right] 
\end{eqnarray}}

In addition to the $CP-$violating contribution due to the mixing of 
the states $|N_i>$ and $|N_i^c>$, which we call $\delta$,
there is another contribution $\epsilon^\prime$ coming from
the direct $CP-$violation through the decays of $|N_i>$ and $|N_i^c>$,
\begin{eqnarray}
\Gamma_{N_i} & = & \frac{1}{2} (1 + \epsilon^\prime) \frac{1}{16 \pi}
               \sum_{\alpha}  |h_{\alpha i}|^2 M_i \nonumber \\
\Gamma_{N_i^c} & = & \frac{1}{2} (1 - \epsilon^\prime) \frac{1}{16 \pi}
               \sum_{\alpha}  |h_{\alpha i}|^2 M_i 
\end{eqnarray}
which has been discussed in the literature extensively \cite{fy,covi}.

Then it is straightforward to show that the asymmetry parameter consists
of the following two parts,
\begin{equation}
\Delta = \epsilon^\prime + \delta
\end{equation}
The new indirect $CP-$violation $\delta$, which enters through the
mass matrix, is given by,

\noindent
\vbox{\begin{eqnarray}
\delta & = & {\rm Re} \left[\sum_{\alpha} h_{\alpha 1}^\ast h_{\alpha 2}
         (\alpha_2 - \alpha_1^\ast)\right] \left(
         \frac{1}{\sum_{\alpha} |h_{\alpha 1}|^2}
         + \frac{1}{\sum_{\alpha} |h_{\alpha 2}|^2 } \right) \nonumber \\
&=& 2 \, \pi \, g^{ab} {\cal C} \frac{M_1 M_2}{M_2^2 - M_1^2} 
\end{eqnarray}}
where,
\begin{equation}
{\cal C} = - {1 \over \pi}
  {\rm Im}[ \sum_\alpha (h_{\alpha 1}^\ast h_{\alpha 2})
 \sum_\beta h_{\beta 1}^\ast h_{\beta 2})] \left( 
         \frac{1}{\sum_{\alpha} |h_{\alpha 1}|^2}
         + \frac{1}{\sum_{\alpha} |h_{\alpha 2}|^2 } \right)
\end{equation}
From this expression it is clear that this contribution becomes 
significant when the two mass eigenvalues are close to each other 
(figure 3).
\begin{figure}
\mbox{}
\vskip 5.25in\relax\noindent\hskip -.8in\relax
\includegraphics{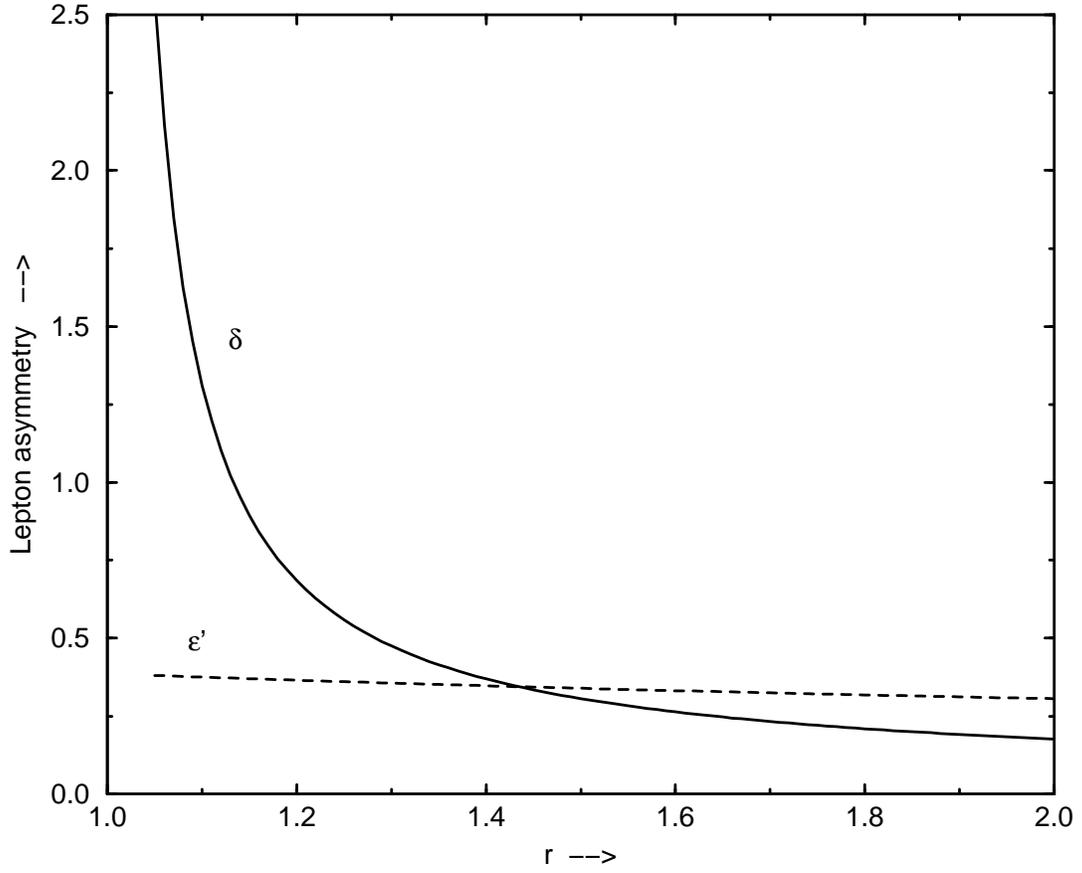}
\caption{Comparison of the two $CP-$violating contributions to the 
lepton asymmetry in units of {\cal C}, where $r = M_2^2/M_1^2$}
\end{figure}
On the other hand the perturbation theory used for 
this expression is valid only for $|M_1 - M_2|
 \gg |H_{ij}^{(1)}| $ or $|\widetilde{H_{ij}^{(1)}}|$.

To find out the value of $\delta$ in the vicinity of $M_1=M_2$ we now 
write  $M_1 = M$ and $M_2 = M + \eta M$ and consider the case 
$M \eta \leq |H_{ij}|$ or equivalently $\eta \leq |\sum_\alpha 
h_{\alpha i} h_{\alpha j}^*|$. Then the total Hamiltonian,
\begin{equation}
\widehat{{\cal H}} = \left( \begin{array}{cccccc}
                  0 & 0 & M + H_{11} & H_{12} \\
                  0 & 0 & H_{12} & (M + \eta M) + H_{22} \\
                  M + H_{11} & H_{12} & 0 & 0 \\
                  H_{12} & (M + \eta M) + H_{22} & 0 & 0 \\
              \end{array}    \right)
\end{equation}
will have the eigenvalues,
\begin{eqnarray}
\Lambda^2 &\approx & M^2(1 + \eta) + M \left\{ H_{11} + H_{22} 
\pm c \right\} \nonumber \\
{\rm or} \hskip .5in \Lambda &\approx & 
\pm \left[ (M + \frac{1}{2} \eta M) + \frac{1}{2} 
          \left\{ H_{11} + H_{22} \pm c \right\} \right]
\end{eqnarray}
with $ c = \sqrt{\left[ M \eta - (H_{11} - H_{22}) \right]^2
          + (H_{12} + \widetilde{H}_{12})^2 }, $ neglecting terms of
order $M^2 |\sum_\alpha 
h_{\alpha i} h_{\alpha j}^*|^3$. The eigenvectors are now given by,
\begin{eqnarray}
|\Psi_1> & = & \frac{1}{\sqrt{2(1 + |\alpha|^2)}} (|N_1> + \alpha |N_2> + |N_1^c>  
               + \alpha |N_2^c>) \nonumber \\
|\Psi_2> & = & \frac{1}{\sqrt{2(1 + |\beta|^2)}} (|N_2> + \beta |N_1> 
               + |N_2^c> +\beta |N_1^c>) 
\end{eqnarray} 
with
$$\alpha = \frac{H_{12} + \widetilde{H}_{12}}{H_{11} - H_{22} - \eta M
         + c} \quad , \quad
\beta = \frac{H_{11} - H_{22} - \eta M - c}{H_{12} + \widetilde{H}_{12}} ,$$

Note that the states $|\psi_1>$ and $|\psi_2>$ are not $CP-$symmetric,
because under $CP-$transformations $\alpha |N_i>$ transforms into 
$\alpha^\ast |N_i^c>$.
With this we can now calculate the lepton asymmetry for the 
degenerate scenario defining the asymmetry parameter as before.
It is now given by,
\begin{equation}
\delta =  {\cal C} 
    \frac{\pi \, g^{ab} \eta}{\eta^2 + (g^{ab})^2 {\rm Re}^2(\sum_\alpha
         h_{\alpha 1}^{\ast} h_{\alpha 2})}
\end{equation}
This expression vanishes for $\eta \to 0$ as expected,
because there is no detectable mixing between two identical particles
(figure 4). 
\begin{figure}
\mbox{}
\vskip 5.75in\relax\noindent\hskip -.8in\relax
\includegraphics{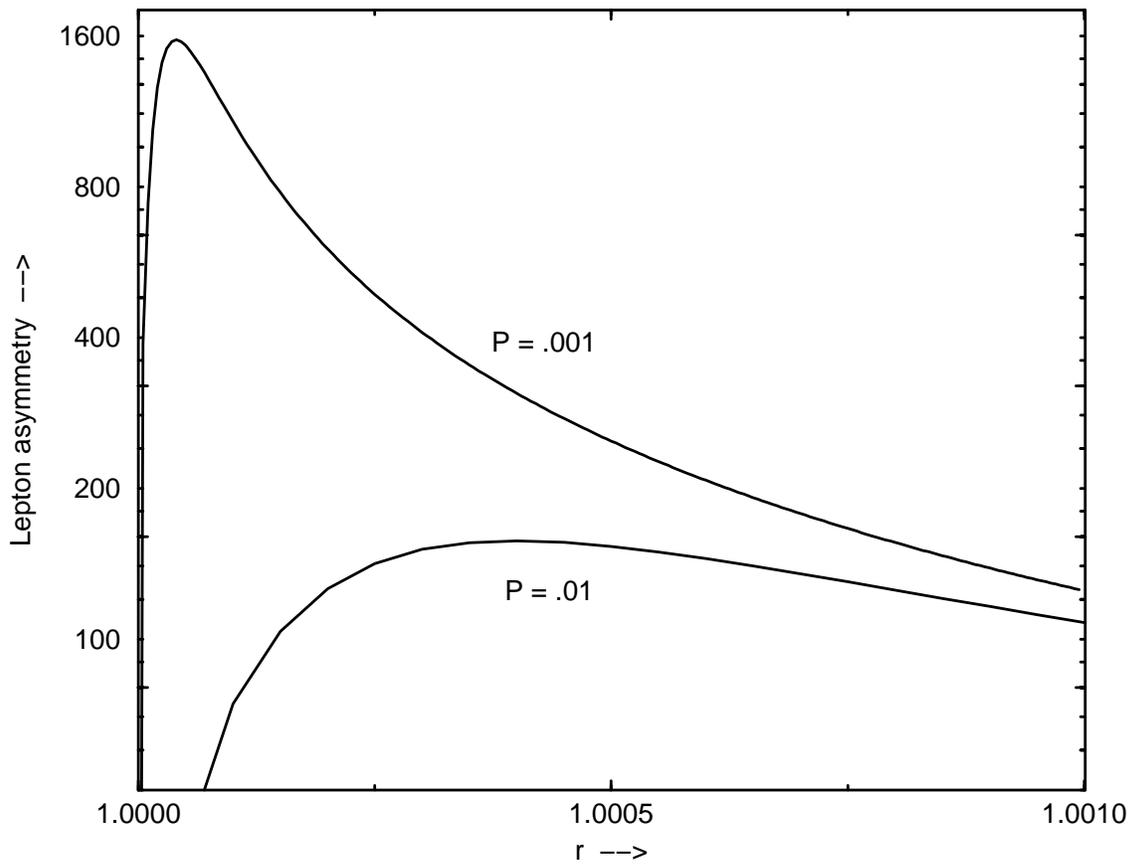}
\caption{Lepton asymmetry in units of ${\cal C}$ generated for a small 
mass difference, where ${\rm P} = \sum_\alpha 
h^\ast_{\alpha 1} h_{\alpha 1} $ and ${\rm r} = M_2^2 / M_1^2$. }
\end{figure}
The matching of the two solutions occurs at $\eta \approx 
{\sum_\alpha h_{\alpha 1}^\ast h_{\alpha 2}}/{5}$
and thus we obtained an asymmetry for small and large mass differences. 
This is shown in figures 3 and 4.
For $\eta \approx |\sum_\alpha h_{\alpha i} h_{\alpha j}^*|$ a big
enhancement of the asymmetry parameter $\delta$, by several orders
of magnitude, is achieved. The enhancement factor is roughly 
${1 / \eta}$. For large values of $\eta$ the two distinct 
contributions $\epsilon^\prime$ and $\delta$ become of the same order.

The  dynamical  evolution of the lepton number in the universe is
governed  by the  Boltzmann  equations.  This is the  appropriate
tool to describe  any  deviation  from thermal  equilibrium.  The
framework and the derivation of the Boltzmann equation is reviewed
in \cite{kolb};  we  adopt  the  same   notation.  We  make  the
approximations  of  kinetic   equilibrium  and   Maxwell-Boltzmann
statistics.  The   out-of-equilibrium   conditions  occur,
when the  temperature  $T$ drops below the
mass  scale  $M_{\psi_1}  = M_1 +  H_{11}^{(1)}  $, where  the
inverse decay is effectively frozen out.

The density of the lepton number  asymmetry $n_L = n_l - n_{l^c}$
for the  left-handed  leptons  can be shown to evolve in time
according to the equation:
\begin{equation}
\frac{{\rm d}n_L}{{\rm d}t} + 3 H n_L = (\epsilon^\prime + \delta_1)
\Gamma_{\psi_1}^{th} [n_{\psi_1} - n_{\psi_1}^{eq} ] - \left(
\frac{n_L}{n_\gamma} \right) n_{\psi_1}^{eq} \Gamma_{\psi_1}^{th} 
-2 n_\gamma n_L \langle \sigma |v| \rangle
\end{equation}
where $\delta_1$ is the contribution proportional to $1/\sum_\alpha
|h_{\alpha 1}|^2$ in the expression for $\delta$.
The second term on the left side comes from the  expansion of the
universe,  where  $H$  is  the  Hubble   constant.  $\Gamma_{\psi
1}^{th}$ is the  thermally  averaged  decay rate of the  $|\psi_1
\rangle$  state,  $n_\gamma$ is the usual photon  density and the
term  $\langle  \sigma  |v|  \rangle$   describes  the  thermally
averaged cross-section of $l + \phi^\dagger \longleftrightarrow l^c +
\phi$  scattering.  We note that the first term of the right side
of this  equation  describes the creation of lepton number and is
proportional to  $(\epsilon^\prime  + \delta_1) $, while the last
two terms are  responsible for any depletion of lepton number and
are   coming   from  the   inverse   decay  and  the   scattering
respectively.  The density of the $\psi_1$  state  satisfies the
Boltzmann equation,
\begin{equation}
\frac{{\rm d}n_{\psi_1}}{{\rm d}t} + 3 H n_{\psi_1} = - 
\Gamma_{\psi_1}^{th} (n_{\psi_1} - n_{\psi_1}^{eq})
\end{equation}
In order to find a solution  to this set of coupled  differential
equations it turns out to be useful to transform to new
variables.  We  introduce   the   dimensionless   variable  $x  =
{M_{\psi_1}}/{T}$,  a particle density per entropy density $Y_i
= {n_i}/{s}$  and make use of the relation $t = {x^2}/{ 2
H (x = 1) }$.

In addition we define the parameter $K = {\Gamma_i (x = 1)}/{H(x = 1)}$
which is a measure of the deviation from equilibrium. For $K \ll 1$
at $T \approx M_{\psi_1}$ we are far from thermal equilibrium so that both
inverse decays and 2 $\leftrightarrow$ 2 $CP$ non-conserving scattering processes
are not important and can be safely ignored. With these simplifications
and the above redefinitions the Boltzmann equations effectively read :
\begin{eqnarray}
\frac{{\rm d}Y_L}{{\rm d}x} & = & (Y_{\psi_1} - Y_{\psi_1}^{eq})
(\epsilon^\prime + \delta_1) K x^2  \nonumber \\
\frac{{\rm d}Y_{\psi_1}}{{\rm d}x} & = & - (Y_{\psi_1} - Y_{\psi_1}^{eq})
K x^2  
\end{eqnarray}
For very large times the solution for $Y_L$ has an asymptotic value
which is approximately given by 
\begin{equation}
Y_L = \frac{n_L}{s} = {1 \over g_*} (\epsilon^\prime + \delta_1)
\end{equation}
where  $g_*$  denotes the total  number of  effectively  massless
degrees of freedom and is of the order of $O(10^2)$ for all usual
extensions  of the  standard  model.  The lepton  asymmetry  thus
generated  will then be converted to the baryon  asymmetry of the
universe during the electroweak phase transition \cite{ht} and is
approximately given by $n_B \approx {1 \over 3} n_L$.  So we have
demonstrated   that  the  baryon  asymmetry  is  proportional  to
$(\epsilon^\prime  + \delta_1)$.  This new contribution was shown
to be at least of the same  order  as the  $\epsilon^\prime$  (as
shown in figure 3) and for small values of r, $\delta$  exceeds
$\epsilon^\prime$ by several orders of magnitude (figure 4).

\vskip .5in

{\bf   Acknowledgement   }  We   thank   Dr.  A.  Pilaftsis   for
discussions.  The  financial   support  of  BMBF  under  contract
056DO93P(5)  is  greatfully  acknowledged.  One of us (US)  would
like to acknowledge a fellowship  from the Alexander von Humboldt
Foundation and hospitality from the Institut f\"{u}r Physik, Univ
Dortmund  during his research  stay in Germany and (MF) wishes to
thank the Deutsche  Forschungsgemeinschaft  for a scholarship  in
the   Graduiertenkolleg   "Production  and  Decay  of  Elementary
Particles".

\end{document}